\documentclass{aastex}
\usepackage{emulateapj5}

\newcommand{\etal}{et~al.\,}
\newcommand{\Msun}{${\rm M}_\odot$}                  
\newcommand{\drp}{\Delta m_{15}(B)} 
\newcommand{\Hunits}{km s$^{-1}$ Mpc$^{-1}$}

\newcommand{\hof}{H\"oflich}

\newcommand{\eg}{e.\,g.\,}

\newcommand{\SiII}{\ion{Si}{2}}
\newcommand{\TiII}{\ion{Ti}{2}}

\shorttitle{SN Ia Progenitors}
\shortauthors{Howell}

\begin{document}

\title{The Progenitors of Subluminous Type Ia Supernovae}
\author{D. Andrew Howell} 
\affil{Lawrence Berkeley National Laboratory\\ 1 Cyclotron Road, Berkeley, CA, 94720}
\email{DAHowell@lbl.gov}

\begin{abstract}
We find that spectroscopically peculiar subluminous SNe Ia come from an old 
population.  Of the sixteen subluminous SNe Ia known, ten are found in 
E/S0 galaxies, and the remainder are found in early-type spirals.  The 
probability that this is a chance occurrence is only 0.2\%.
The finding that subluminous SNe Ia are associated with an older stellar
population indicates that for a sufficiently large lookback time
(already accessible in current high redshift searches) they will not be 
found.  Due to a scarcity in old populations, hydrogen and helium main 
sequence stars and He red giant stars that undergo Roche lobe overflow 
are unlikely to be the progenitors of subluminous SNe Ia.
Earlier findings that overluminous SNe Ia ($\drp \lesssim 0.95$) 
come from a young progenitor population are confirmed.  The fact that 
subluminous SNe Ia and overluminous SNe Ia come from different progenitor 
populations and also have different properties is a prediction of the CO 
white dwarf merger progenitor scenario. 

\end{abstract}

\keywords{supernovae: general}

\section{Introduction}
Type Ia supernovae (SNe Ia) have been used to measure the age, 
mass density, and dark energy content of the universe (Riess \etal\ 1998, 
Perlmutter \etal\ 1999).  This is possible because they make excellent
calibrated candles --- the shape of their light curves is correlated with
their luminosity.  SNe Ia that slowly decline in brightness (and have
a lower value of the $\drp$ parameter) are more luminous than SNe Ia 
that are fast decliners (Phillips 1993). 

There is near-universal agreement that Type Ia supernovae (SNe Ia) are the 
result of the explosion
of accreting CO white dwarfs in binary systems (Hoyle \& Fowler, 
1960), though the nature of the accretion process and the type of
the secondary star remain unknown.  If the age of Type Ia SNe can be 
identified, then it would limit the mass of the progenitor star and possible
progenitor scenarios. 

SNe Ia show variations in properties with environment, as
noted by \eg\ Filippenko (1989), Branch \& van den Bergh (1993) 
Fisher \etal\ (1995), Hamuy et al. (1996), and Riess \etal\ (1999).  

Many previous authors have attempted to estimate the age of SNe Ia, 
including: Oemler \& Tinsley (1979; $10^7-10^8$ yr), Bartunov \etal\ 
(1994; $\ge 10^7$), McMillan \& Ciardullo (1996; $\geq 5 \times 10^8$ yr),
and Yoshii, Tsujimoto, \& Nomoto (1996; $0.5 - 3 \times 10^9$ yr).

The above studies assumed a single progenitor population
for SNe Ia.  The only way that the presence of SNe Ia in E galaxies can be
reconciled with intermediate-age progenitors is if they stopped star
formation relatively recently.  Alternatively, as suggested by 
Dallaporta (1973) and Della Valle \& Livio (1994) 
there may be two channels to SNe Ia --- from both old and young stellar 
populations.  

Hamuy \etal\ (1995, 1996) noticed that the slowest declining SNe only occur in 
spiral galaxies.  However, they stopped short of claiming that the fast 
declining SNe are associated with an older population, since these SNe 
were seen in both E/S0 galaxies and spirals.  Hamuy \etal\ 2000 also 
examined the relationship of SN Ia luminosity to environment, but came to 
no strong conclusions about subluminous SNe.  In this study we focus on the 
most subluminous, spectroscopically peculiar SNe Ia where any age effects 
should be strongest, then examine other classes of SNe Ia. 

\section{Method}
We compiled a list of all peculiar SNe Ia known at the time of writing.
These SNe are listed in Table~\ref{pec-table} along with their host galaxies 
and types.  The data 
were primarily taken from  Li \etal\ (2001), Branch, Fisher, \& 
Nugent (1993), Phillips \etal\ (1999), the IAU circulars, and the 
Asiago SN Catalog (Barbon \etal\ 1999).  We define SNe Ia 
as subluminous if they show spectroscopic 
signatures of subluminous SNe Ia such as deep, broad, \TiII\ around 
4200 \AA\ combined with enhanced \SiII\ 
5800 \AA\ absorption.  SN 1991bg and SN 1986G showed both of these signatures,
as well as a fast-declining light curve.  In the absence of published
spectroscopic information, one SN in this study, SN 1996bk, was judged to be
subluminous due to the fact that it has a light curve decline rate 
parameter, $\drp=1.75$, greater than that of SN 1986G (1.73).
Unless otherwise stated, all $\drp$ information in this paper is from
Phillips \etal\ (1999).

Overluminous SNe are defined as SNe that are spectroscopically similar
to SN 1999aa or SN 1991T (see \eg\ Li \etal\ 2001), or those with a slowly 
declining light curve ($\drp \leq 0.95$).  These SNe show \ion{Fe}{3}, and  
\ion{Si}{2} is weak or absent.  SN 1999aa shows \ion{Ca}{2}, whereas SN 1991T
does not.  We note that photometry
has not yet been published for SN 1999aa and that it may not turn out
to be overluminous.  We still retain this label for these SNe which 
are spectroscopically similar to SN 1991T.  For this study, most SNe were
classified based on spectroscopic peculiarity, not photometric light 
curves.

\section{Results}
\subsection{Subluminous SNe Ia} 
From the collection of subluminous SNe Ia, it is apparent that they favor 
E and S0 galaxies.  Ten of the sixteen subluminous SNe Ia come from 
either E or S0 (E/S0) host galaxies.  To calculate the probability that
this is a coincidence, first we must know the chance that a random SN's 
host galaxy will be an E or S0 galaxy.  Barbon \etal\ (1999) gives host 
galaxy statistics from the Asiago SN catalog.  Out of 235 SNe Ia 
(including peculiars), with known galaxy types 62 (26\%) were from 
E/S0 galaxies. 

We can estimate the probability that at least 10 of 16 randomly drawn SNe Ia
would be from E/S0 galaxies using the binomial distribution (Bernoulli 1713).
If $\pi$ is the probability that the host galaxy for a given SN Ia is
an E or S0 galaxy, and we select $N$ random SNe Ia, then the probability
that at least $r$ host galaxies will be of type E or S0 is
$$\sum_{i=r}^{N}P(r_i)=\frac{N!}{r_i!(N-r_i)!}\pi^{r_i}(1-\pi)^{N-r_i}$$
For $\pi=0.26$, $N=16$, and $r=10$, $P=0.2\%$.  
Thus the probability that it is a coincidence that 10 of the 16 
subluminous SNe Ia are from E or S0 galaxies is 0.2\%.  

Two SNe Ia cannot be identified with certainty as subluminous due to a
lack of data, but are suspicious: SN 1980I, and SN 1971I 
(Branch \etal\ 1993).  If we include these uncertain events as subluminous 
SNe Ia, then 11 of 18 are in E/S0 galaxies and $P$ is still 0.2\%.

These data indicate that the association of subluminous SNe Ia with 
early-type galaxies is real.  Furthermore, it is evidence that 
subluminous SNe Ia come from an older stellar population.
This is a subtle point that deserves explanation.  In the larger
sample of SNe Ia as a whole, E and S0 galaxies make up only a small
fraction (26\%) of the hosts.  As long as these hosts are a minority,
one may speculate that SNe Ia come from an intermediate-age population, 
and SNe Ia only happen in E or S0 galaxies with recent star formation
activity.  However for a subsample of SNe Ia that greatly favors 
E or S0 hosts, these SNe Ia cannot be considered an oddity, but
must be associated with some intrinsic feature of the host galaxy type.  
In other words, the property of E and S0 galaxies that makes them 
unique, an older average stellar population, must be the culprit 
(metallicity is likely a second-order effect, see, \eg\ 
\hof\ \etal\ 1998, 2000; Umeda \etal\ 1999; Ivanov \etal\ 2000).

Using line strength indicators, Trager \etal\ determine the age of NGC 4374, 
the host galaxy of the strongly subluminous SN 1991bg, to be approximately 
$14\pm2.5$ Gyr.  This age relies on the assumption of a single
burst of star formation, and is model dependent, but it seems a safe bet 
that star formation stopped in NGC 4374 at least 10 Gyr ago, giving us a 
rough lower limit for the age of the progenitor of SN 1991bg.

\subsection{Overluminous SNe Ia}
Several previous authors have noted that overluminous SNe Ia, such as
SN 1991T, tend to favor star forming environments and may be associated
with a younger stellar population (\eg\ Hamuy \etal\ 1996).  These suspicions 
are upheld in the data presented here.  

From the Asiago catalog, 42\% of SNe Ia are from Sbc or later host 
galaxies.  In the sample presented here, if we consider all events (even those
with uncertainties), then 17 of 23 overluminous SNe Ia are from this late 
set of hosts.  The binomial distribution gives a 0.2\% probability 
that this is a chance occurrence.  If we throw out the SNe Ia with 
uncertainties in their classification or host galaxy type classification, 
and also remove SNe Ia in peculiar galaxies, then we are left with 14
overluminous SNe Ia.  Of these, 12 are in Sbc or later galaxies.  There is
only a 0.1\% probability that this happened by chance.  

While no overluminous SNe Ia have been discovered in E galaxies, a few have
been discovered in S0 galaxies: SNe 1998es, 1998ci, and 2000cx.
Of these galaxies, the host of SN 1998ci is listed as peculiar, and inspection
of CCD images reveals that this galaxy may have a distorted disk.  
Additionally, the host galaxy of SN 1998es, NGC 632, has evidence for a 
nuclear starburst and its type is uncertain (Pogge \& Eskridge 1993).
SN 2000cx has a more normal S0 host, but it is not a typical SN 1991T-like
event since it shows a peculiar spectral evolution (Weidong Li, private 
communication).

There is an additional piece of evidence that SN 1991T-like objects 
are found in a young stellar population --- the presence of circumstellar
material.  Five supernovae have been spectroscopically confirmed as true
SN 1991T-like events (not SN 1999aa-like): SN 1991T, SN 1995ac, SN 1995bd, 
SN 1998ci, and SN 1999J.  Of these, {\em all} five show signs of reddening. 

\subsection{Selection Effects}
The SNe studied in this paper were culled from a diverse array of
sources, so selection effects may not be negligible.  To estimate the role of 
selection effects we also test SNe from a more uniform sample.  
Li \etal\ (2001b) present a sample of local SNe Ia discovered
by the Beijing Astronomical Observatory Supernova Search (BAOSS) and
the Lick Observatory Supernova Search (LOSS).  These CCD searches were done
using similar techniques and attempt to avoid many of the pitfalls of
magnitude limited searches.  This sample of SNe should be much less 
affected by selection effects than a random collection.

It is apparent that subluminous SNe Ia are preferentially located in E or S0 
galaxies in this sample as well.  Five of seven subluminous SNe Ia are in 
E/SO galaxies, giving $P=0.5\%$.  Selection effects do not seem to be the
culprit in the tendency for subluminous SNe Ia to favor older stellar 
populations.

\subsection{Diversity in the Older Population}
\begin{figure}[!htp]
\plotone{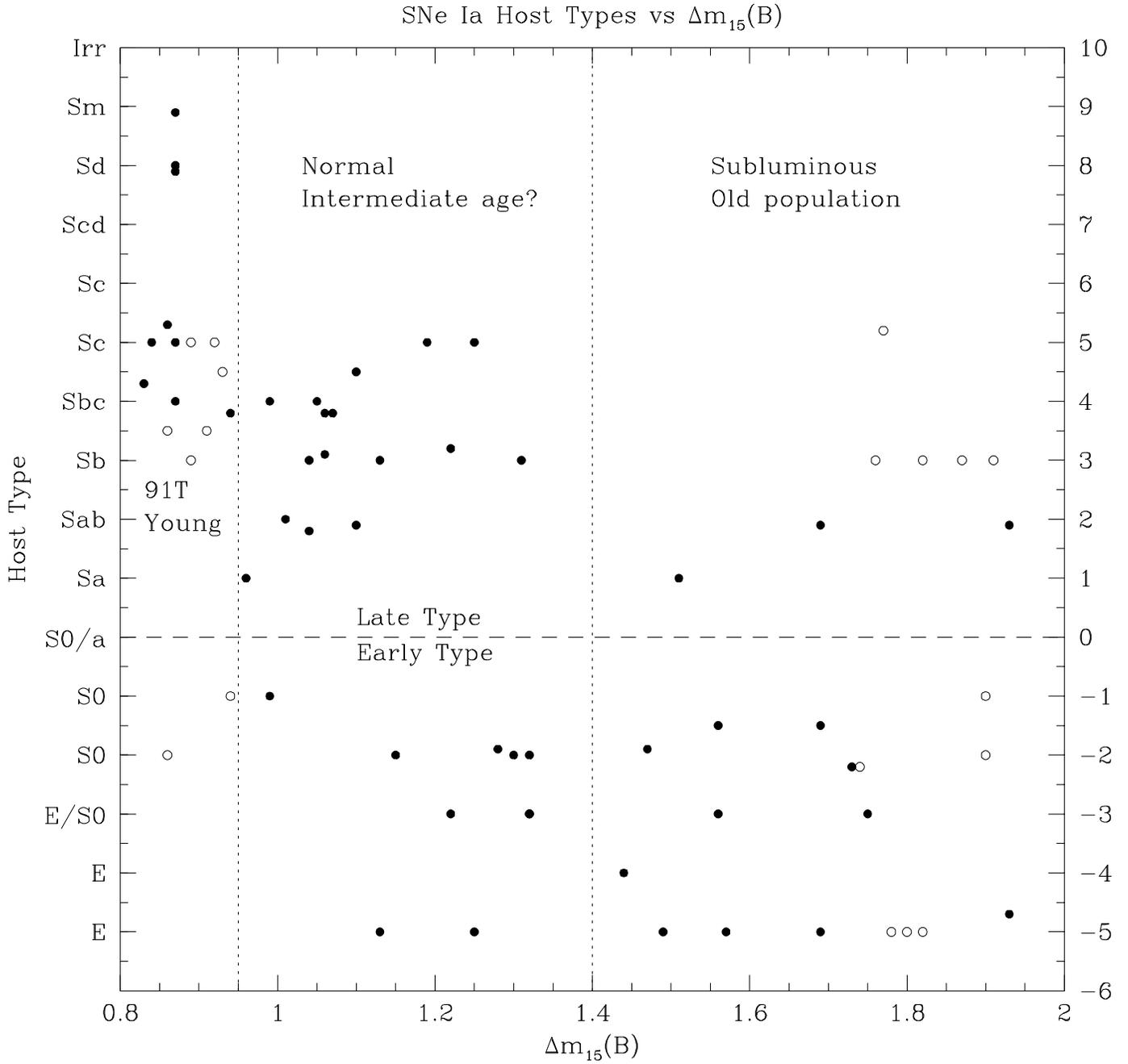}
\caption{SNe Ia host galaxy types vs $\drp$.  Closed circles are 
photometrically determined.  Open circles are spectroscopically peculiar 
SNe Ia for which photometric decline rates are not available.  They are 
randomly assigned deline rates within the ranges of
SNe Ia that are spectroscopically similar ($\drp \ge 1.73$ for 
strictly-overluminous SNe and $\drp \le 0.95$ for overluminous SNe).  
Note that this assumes that decline rate and spectral type are well 
correlated (Nugent \etal\ 1995).  Only SNe Ia with well-determined host 
galaxy types are plotted.  SNe are plotted against T type (right axis), 
and the left axis just exists as a guide, so some galaxy types are 
listed twice.}  
\label{drp-fig}
\end{figure}

Figure~\ref{drp-fig} shows SNe Ia with well-determined, reddening-corrected
decline rates.  Host galaxy types were taken from the Asiago SN Catalog. 

From this figure, it appears that the sample can be divided up into
three regions, SNe Ia with $\drp \le 0.95$ (overluminous), those with
$0.95 < \drp < 1.4$ (normal) and those with $\drp \ge 1.4$ 
(loosely-subluminous).  We note that the division at $\drp = 1.4$
may not be exact.  The line could be drawn at $\drp=1.3$, for example.
We designate SNe Ia with $\drp \ge 1.4$ as 
loosely-subluminous to differentiate them from spectroscopically peculiar,
very subluminous SNe Ia like SN 1991bg and what may be its weaker cousin 
SN 1986G.  We will now call these latter SNe Ia strictly-subluminous.  
We take the dividing line of strictly-subluminous SNe Ia at the value 
for SN 1986G, $\drp \ge 1.73$.

A striking feature about Figure~\ref{drp-fig} is that strictly-subluminous
SNe Ia are not the only ones that show a preference for E/S0 galaxies.
For SNe with $\drp \ge 1.4$ (and spectroscopically confirmed subluminous
SNe Ia) 18 of 26 in this sample were in E/S0 host galaxies.  The probability 
that this is
a chance occurrence is $5 \times 10^{-6}$.  This is interesting because SNe Ia with
$\drp \sim 1.4$ are essentially spectroscopically normal.  
This indicates that a dramatically different mechanism may not be at
work in the normal and loosely-subluminous SNe Ia.  

Overluminous SNe Ia appear to come exclusively from a younger population,
while subluminous SNe Ia come from an old population.  This raises an 
obvious question:  Is the age of the progenitor star the parameter that
controls the strength of the explosion?  
Despite the fact that old progenitors appear to be required to produce
subluminous SNe Ia, they also produce fairly normal SNe Ia.  
Figure~\ref{drp-fig} shows that SNe Ia in E 
galaxies have $\drp$ values stretching from 1.13 to 1.93. 
Additionaly, the 10 Gyr old population of NGC 4374 that was host to SN 1991bg 
was also the home of SN 1957B.  This SN was subluminous by about 0.9 mag in 
$B$ relative to a typical SN Ia (Della Valle \etal\ 1998), but brighter than 
SN 1991bg.  It had a decline rate of 1.3 mag.  

\section{Conclusions}
Earlier findings that overluminous SNe Ia come from a young progenitor 
population were confirmed, as 17 of 23 are from Sbc or later host galaxy 
types.  If we disregard SNe with peculiar and uncertain host galaxy types and 
SNe whose classifications are in doubt, then a remarkable 12 or 14 
overluminous SNe are from Sbc or later galaxies.

We also find that spectroscopically subluminous SNe Ia come from an old 
population, in the case of SN 1991bg, at least 10 Gyr old.  The simplest 
hypothesis is that all subluminous SNe Ia are from $\sim$ 10 Gyr old 
progenitors, but we cannot rule out a spread of ages between $\sim$ 
5 -- 12 Gyr.  More mildly subluminous SNe Ia ($\drp \gtrsim 1.4$) also show 
a preference for early-type galaxies.  This is consistent with the findings 
of Ivanov \etal\ 2000.

If the progenitors of subluminous SNe Ia are $\sim$ 10 Gyr old, then they
should not exist at high redshift.  For $\Omega_M=0.3$, 
$\Omega_{\Lambda}=0.7$, and $H_0=65$ \Hunits , at $z=0.5$, where the bulk
of SNe Ia in current studies are clustered, the age of the 
universe is 9.1 Gyr.  Indeed, no
subluminous SNe Ia are seen at $z>0.4$ in the studies of 
Riess \etal\ (1998) or Perlmutter \etal\ (1999).  Due to selection effects, 
however, we cannot conclude that the lack of subluminous SNe Ia at high 
redshift is due to evolution.  

The finding that at least some subluminous SNe Ia must come from very
old progenitors places tight constraints on their progenitor systems.
Only low mass ($\lesssim$ 1 \Msun ) nondegenerate stars will exist in a 10 Gyr
old stellar population.   It unlikely that the following scenarios could 
produce SNe Ia in such an old population:  hydrogen (H) main sequence (MS) 
stars that undergo Roche lobe overflow (RLOF) or helium (He) MS or 
red giant (RG) stars that undergo RLOF (see Branch \etal\ 1995 and references
therein, hereafter B95).  

The remaining possible scenarios for subluminous SNe are hydrogen
RG stars with RLOF, H stars that lose mass through a stellar wind 
(symbiotic stars), sub-Chandrasekhar mass explosions, and WD mergers.
Sub-Chandrasekhar mass models are ruled out because they do not 
match observations (H\"oflich et al. 1996; Nugent et al. 1997).
Of the remaining scenarios, each has problems.  H RG RLOF systems are 
not thought to exist in great numbers (B95).  Symbiotic stars may not exist in 
significant numbers in such old populations (B95), may not be accrete enough 
hydrogen, and should produce interaction with the circumstellar medium 
with has not been observed (Eck \etal\ 1995).  
Neither of the latter two cases should produce a significant number of 
SNe in young populations, meaning that if slow accretion 
to near the Chandrasekhar mass is the mechanism behind a SN Ia, then
separate channels are required to produce subluminous and overluminous SNe Ia.

The finding that multiple channels are required to explain the age 
difference between SNe Ia is contrary to the suggestion of Branch (2001) 
that multiple channels are unnecessary.  If there is indeed only a single
channel to SNe Ia, then it must be the double-degenerate 
(merging WD) scenario (Webbink 1984, Iben \& Tutukov 1984).  In this 
scenario two CO white dwarfs merge to produce a Chandrasekhar mass 
(or greater) SN.  
One of the most appealing aspects of this theory is that more luminous 
explosions are naturally expected in younger systems (Tutukov \& Yungelson 
1994, 1996).  This is because in younger systems more massive stars are 
evolving to more massive WDs than in older systems.  For $t \lesssim 10^{8.5}$
yr, the average total mass of the system is expected to be 2.1 \Msun .  
SNe Ia arising from older systems would be less massive.  Mergers would also 
be attractive for producing the asymmetries seen in SN 1999by 
(Howell \etal\ 2001), but some doubts exist as to whether or not they
produce SNe Ia (\eg\ Saio \& Nomoto 1985).

\acknowledgements
The author thanks J. Craig Wheeler, Peter \hof , Lifan Wang, Peter 
Nugent, and Alex Filippenko for helpful comments and discussion, as well 
as Weidong Li for providing information prior to publication.   

This work was done primarily at the University of Texas in partial
fulfillment of the requirements of a PhD thesis (Howell 2000), supported by 
grant 9818960 from the National Science Foundation.  
This work was also supported in part by the Physics Division, E. O. Lawrence 
Berkeley National Laboratory of the U.S. Department of Energy under Contract 
DE-AC03-76SF000098.  This research also made use of the NASA/IPAC
Extragalactic Database (NED), which is operated by the Jet Propulsion
Laboratory, California Institute of Technology, under contract with
NASA.  It also made use of the list of supernovae provided by
the Central Bureau for Astronomical Telegrams operated at the 
Smithsonian Astrophysical Observatory, under the auspices of 
Commission 6 of the International Astronomical Union.

\begin{deluxetable}{llllc}
\tabletypesize{\scriptsize}
\tablewidth{0pt}
\tablecaption{Peculiar SNe Ia}
\tablehead{
\colhead{SN } &\colhead{T\tablenotemark{a}}& \colhead{Galaxy} & \colhead{G type} & \colhead{Ref}
}
\startdata
1957A  &u            &  NGC 2841     &Sb     &1\nl                        
1960H  &u            &  NGC 4096     &Sc    &1\nl 			     
1971I  &u:           &  NGC 5055     &Sbc&1\nl			      
1980I  &u:           &  NGC 4387:\tablenotemark{b}    &E      &2\nl 
1986G  &u            &  NGC 5128     &S0     &3\nl                        
1991F  &u            &  NGC 3458     &S0     &4\nl  			     
1991bg &u            &  NGC 4374     &E      &3\nl  			     
1992K  &u            &  E269-G57     &Sab   &3\nl			      
1993R  &u            &  NGC 7742     &Sb     &5\nl
1996bk &u\tablenotemark{c}  &  NGC 5308     &S0     &6\nl     
1997cn &u            &  NGC 5490     &E      &7\nl  			      
1998bp &u     	     &  NGC 6495     &E	     &8\nl 		      
1998de &u            &  NGC 252      &S0     &7\nl  			      
1999bh &u            &  NGC 3435     &Sb     &8\nl		      
1999by &u            &  NGC 2841     &Sb     &7\nl			      
1999da &u            &  NGC 6411     &E	     &8\nl 		      
1999dg &u            &  UGC 9758     &S0     &8\nl			      
2000ej &u            &  IC 1371      &S0- pec &9\nl
1937C  &o\tablenotemark{c}  &  IC 4182      &Sm     &10\nl      
1972E  &o\tablenotemark{c}  &  NGC 5253     &Sd     &10\nl       
1988G  &o:           &  Anon.        &Sc &1\nl                        
1991T  &o            &  NGC 4527     &Sbc   &3\nl 			      
1991ag &o\tablenotemark{c}  &  IC 4919      &Sd    &10\nl       
1992P  &o\tablenotemark{c}  &  IC 3690      &Sbc    &10\nl            
1992bc &o\tablenotemark{c}  &  E300-G09     &Sc     &10\nl    
1994ae &o\tablenotemark{c}  &  NGC 3370     &Sc     &10\nl            
1996bv &o\tablenotemark{c}        &  UGC 3432     &Scd:   &10\nl      
1995ac &o            &  Anon.        &S      &3\nl     
1995al &o\tablenotemark{c}        &  NGC 3021     &Sbc    &10\nl        
1995bd &o            &  UGC 3151     &Sc     &11\nl                  
1997br &o            &  E576-G40     &Sd:pec&7\nl         
1997cw &o            &  NGC 105      &Sab:   &7\nl         
1998ab &o            &  NGC 4704     &Sbcpec&7\nl           
1998es &o            &  NGC 632      &S0:    &7\nl                 
1998ci &o            &  Anon.        &S0+ pec:&12\nl
1999J  &o            &  Anon.        &?       &13\nl                 
1999aa &o            &  NGC 2595     &Sc     &7\nl             
1999ac &o            &  NGC 6063     &Scd:   &7\nl        
1999cw &o            &  MCG01-02-0   &Sab    &8\nl                    
1999dq &o            &  NGC 976      &Sc     &7\nl                     
1999gp &o            &  UGC 1993     &Sb     &8\nl   
2000cx &o:\tablenotemark{d} &  NGC 524      &S0     &14\nl                 
2001ah &o            &  UGC 6211     &Sbc    &15\nl

\enddata    
\tablenotetext{a} {Whether the SN is like SN 1991bg or SN 1986G and is thus 
underluminous (u) or like SN 1999aa or SN 1991T and is thus overluminous (o).  
A colon denotes uncertainty.}  
\tablenotetext{b}{SN 1980I was between NGCs 4387, 4374, and 4406.  All are elliptical galaxies.}
\tablenotetext{c} {Photometric information was primarily used to make the 
classification.  In all other cases spectroscopic information was used.}
\tablenotetext{d} {SN 2000cx is included for completeness, though it is
not a typical 91T-like event (Weidong Li, private communication).} 
\tablerefs{
(1) Branch, Fisher \& Nugent 1993 
(2) Smith 1981 
(3) Asiago 
(4) Gomez \& Lopez 1995 
(5) IAUC 5842 
(6) Riess \etal\ 1999 
(7) Both Li and Asiago 
(8) Li \etal\ 2000 
(9) IAUC 7532 
(10) Phillips \etal\ 1999 
(11) IAUC 6278 
(12) IAUC 6921 and Nugent, private communication 
(13) IAUC 7096 
(14) IAUC 7463 
(15) IAUC 7604} 
\label{pec-table}
\end{deluxetable}

\clearpage

\end{document}